\documentclass[%
 twocolumn,
 amsmath,amssymb,
]{revtex4-1}

\usepackage{graphicx}
\usepackage{dcolumn,color}
\usepackage{bm}

\usepackage{graphicx}

\input{epsf}

\providecommand\bnabla{\mathbf{\nabla}}

\newcommand\p{\ensuremath{\partial}}

\newcommand{\vetu}[0]{\mbox{\bf{\em{u}}}}

\newcommand{\vete}[0]{\mbox{\bf{\em{e}}}}

\newlength{\fwidth}

\begin{document}



%
\title{Recurrent bursts via linear processes in turbulent environments}
%

\author{Geert Brethouwer$^1$}
 \email{geert@mech.kth.se}
\author{Philipp Schlatter$^1$}%
\author{Yohann Duguet$^2$}
\author{Dan S. Henningson$^1$}%
\author{Arne V. Johansson$^1$}%
\affiliation{%
$^1$Linn{\'e} FLOW Centre and Swedish e-Science Research Centre, KTH Mechanics, SE-100 44 Stockholm, Sweden\\
$^2$LIMSI-CNRS, UPR 3251, Universit\'e Paris-Sud F-91403 Orsay, France
}%


%

\date{\today}

\begin{abstract}
Large-scale instabilities occurring in the presence of small-scale turbulent fluctuations
are frequently observed in geophysical or astrophysical contexts but are difficult to reproduce in the laboratory.
 Using extensive numerical simulations, we report here on intense recurrent 
bursts of turbulence in plane Poiseuille flow rotating about a 
spanwise axis. A simple model based on the linear instability of 
the mean flow can predict the structure and time scale of the 
 nearly-periodic and self-sustained burst 
cycles. Rotating Poiseuille flow is suggested as a prototype for 
future studies of low-dimensional dynamics embedded in 
strongly turbulent environments.

\end{abstract}

\pacs{47.20.-k,47.27.ek,47.27.ed,47.27.nd,47.32.Ef}
\maketitle

Environments with strong fluctuations
frequently display regular or chaotic large-scale dynamics.
Well-known astrophysical examples are the reversals of large-scale planetary magnetic fields,
which are chaotic for the Earth but time-periodic for the Sun \cite{dynamos}.
Other geophysical and astrophysical manifestations of large-scale instabilities include climate cycles on Earth
\cite{climate} and solar flares \cite{solar_flares}.
Random reversals of a large-scale circulation are also found
in Rayleigh-B{\'e}nard convection \cite{rb}, and in von K{\'a}rm{\'a}n \cite{karman}
and laboratory fluid dynamo experiments \cite{riga_vks}.

Recurrent large-scale oscillations or bursts 
like in tokamak plasmas
\cite{tokamaks} and accretion disks \cite{accretion_disks} 
are important manifestations of large-scale instabilities in the presence of (strong) fluctuations. 
Because of their complexity and the very different scales involved, bursting phenomena are frequently
investigated in the laboratory using simplified flow prototypes
that capture the essential features of these large-scale instabilities.
Such prototypes are usually based on simple geometries while the dominant mechanisms under study are preserved, e.g.
the interaction between shear and rotation, convection or Lorentz forces.
A bifurcation from full turbulence to an intermittently bursting
turbulent regime was recently observed in large-gap Taylor-Couette flows with
counter-rotating cylinders \cite{brauckmann}. The bifurcation was found to coincide with the optimal torque parameters.
A striking feature of the large-scale instabilities in the aforementioned systems
is their apparent low-dimensionality, despite the fact that they happen in
environments with intrinsic strong (usually turbulent) fluctuations.
Nevertheless, the fundamental cause of these large-scale dynamics is not always
understood and it is often arduous to derive elementary models from first principles.


In this letter we describe the occurrence of
violent \emph{time-periodic} bursts
in an as yet unexplored parameter range of
turbulent plane rotating Poiseuille flow (RPF),
seen as a canonical example of interaction between shear and rotation. 
The flow develops a large-scale linear instability
under the influence of rotation even though it is strongly turbulent.
This linear instability 
is followed by a distinct
sequence of processes leading to a self-sustaining cycle
of recurrent bursts of turbulence.
%
%
We demonstrate that
the linear instability of the mean flow captures
the essential features of the recurrent bursts. We thus argue that RPF is a relevant example
of low-dimensional dynamics embedded in a high-dimensional system with strong fluctuations, and can serve as a new prototype for future studies of 
large-scale dynamics.

\begin{figure}
\includegraphics[width=60mm]{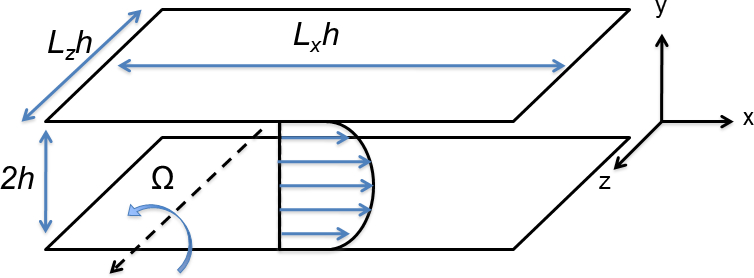}
\caption{Schematic of RPF geometry.}
\label{geo}
\end{figure}

The RPF case considered here is a
pressure-driven plane channel flow
between two smooth parallel flat walls 
subject to a global rotation about the spanwise axis 
orthogonal to the mean flow
and parallel to the walls, see Fig.~\ref{geo} for a schematic.
The velocity field $\vetu$ is governed by the 
incompressible Navier-Stokes equations
in the rotating frame
\begin{eqnarray}
\frac{\p \vetu}{\p t} + \vetu \cdot \bnabla \vetu &=& 
- \bnabla p + \frac{1}{R} \bnabla^2 \vetu - \Omega (\vete_z \times \vetu),\\
\bnabla \cdot \vetu &=& 0,
\end{eqnarray}
where $\vete_z$ is the unit vector in the 
spanwise $z$-direction and $p$ is a modified
pressure including the (dynamically inactive) centrifugal force
\cite{tritton}. 
Streamwise and wall-normal coordinates are 
denoted by $x$ and $y$, respectively.
The equations are nondimensionalized by the
channel half gap $h$ and bulk velocity $\mathcal{U}$ (an average in the $y$-direction),
which is kept constant in time,
giving $R = \mathcal{U} h / \nu$ and $\Omega = 2 \Omega^f h / \mathcal{U}$, 
where $\nu$ is the kinematic viscosity, $\Omega^f$ is the imposed dimensional
global rotation rate, and time $t$ is nondimensionalized with $h/\mathcal{U}$.

The governing equations are rewritten in divergence-free 
velocity-vorticity formulation and projected numerically on a 
Fourier-Chebyshev spectral expansion. Time is advanced using a 
standard semi-implicit Crank-Nicolson/Runge-Kutta scheme \cite{chev}.
Boundary conditions are periodic in $x,z$ and 
no-slip at $y=\pm 1$.
The streamwise and spanwise domain size,
$8 \pi$ and $3 \pi$ respectively,
are large enough to capture large-scale
intermittency and long wavelength instabilities. 
A resolution of $1152 \times 217 \times 864$ spectral 
collocation points in $x,y,z$
is used to fully resolve turbulence.


Rotation about the spanwise axis leads to an
asymmetric mean velocity profile in
three-dimensional turbulent RPF 
since it amplifies
turbulence on the channel side
where the vorticity 
associated with the mean flow $-\vete_z \mbox{d}U/\mbox{d}y$
is antiparallel to the rotation vector $\Omega \vete_z$,
whereas on the other channel side
they are parallel and turbulence is damped \cite{olof}.
Henceforth, we refer to these highly and weaker turbulent channel side
as HTS and WTS, respectively.


RPF at $R=20000$ bifurcates from homogeneous to 
spatially intermittent turbulence 
on the WTS and the mean wall shear stress decreases with
increasing $\Omega$. At $\Omega=0.45$ oblique turbulent-laminar bands
develop. 
Similar intermittent patterns have been 
identified in several transitional flows
\cite{patterns}, but in RPF they are
confined near the wall, as reported in 
other shear flows with damping external forces \cite{brethouwer}.
The second bifurcation from spatially intermittent turbulence to cyclic 
turbulent bursts takes place slightly below $\Omega=0.9$.

\begin{figure}
\setlength{\fwidth}{0.34\linewidth}
\setlength{\unitlength}{\fwidth}
\setlength{\unitlength}{1cm}
\includegraphics[width=50mm]{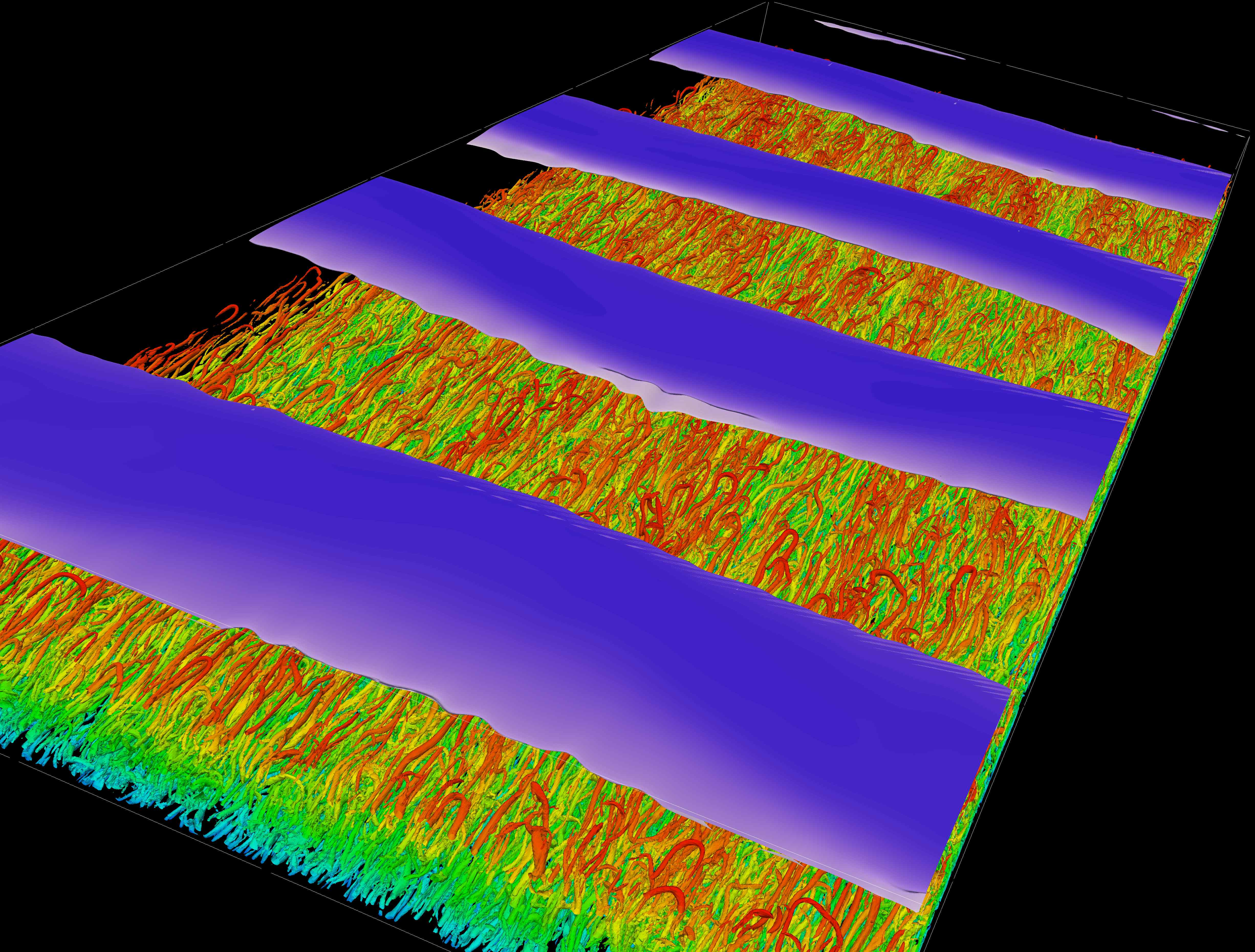}
\put(0.2,3.5){(a)}
\vskip2mm
\includegraphics[width=50mm]{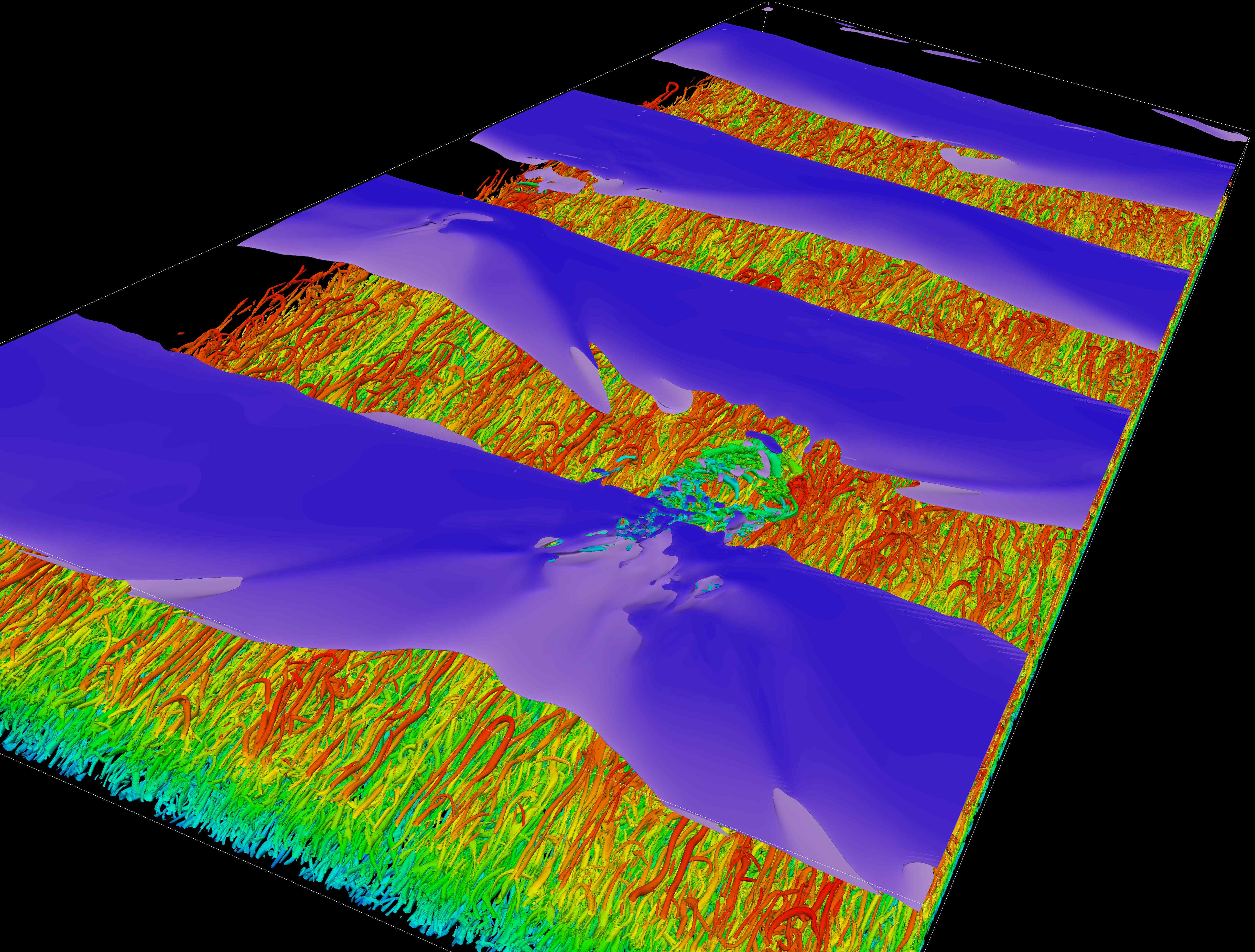}
\put(0.2,3.5){(b)}
\vskip2mm
\includegraphics[width=50mm]{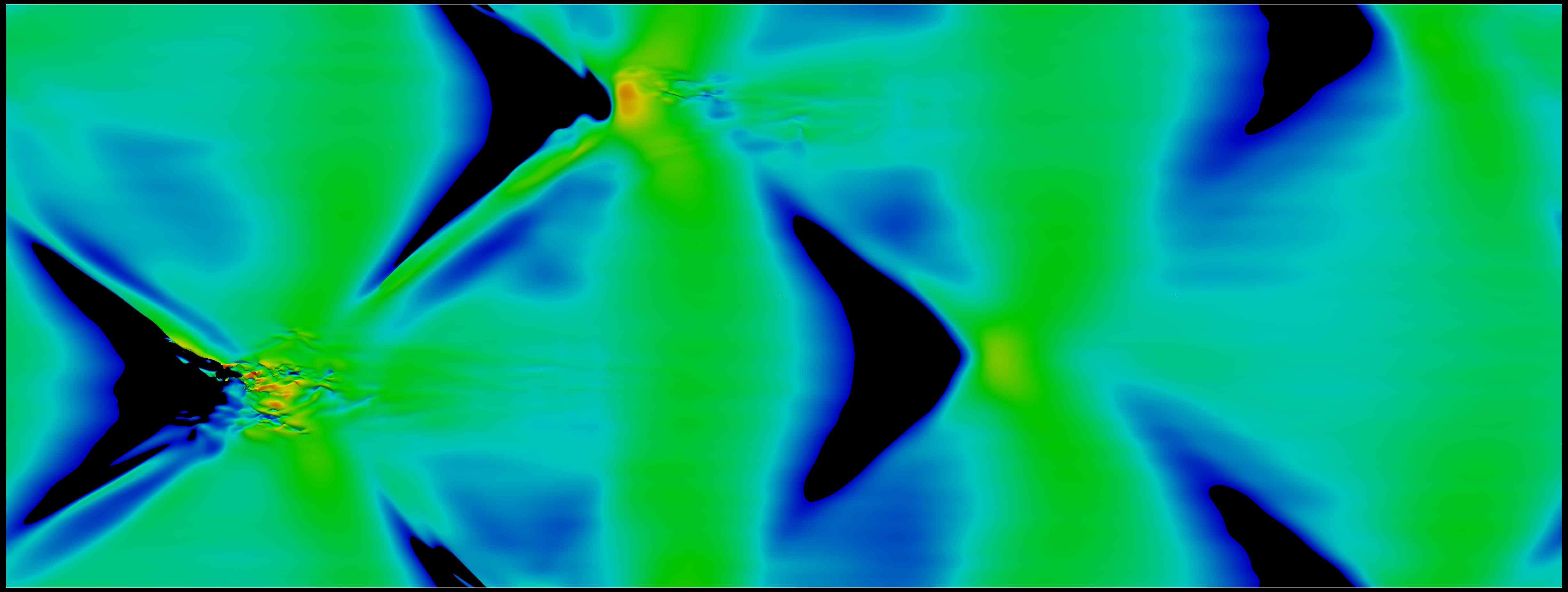}
\put(0.2,1.6){(c)}
\vskip2mm
\includegraphics[width=50mm]{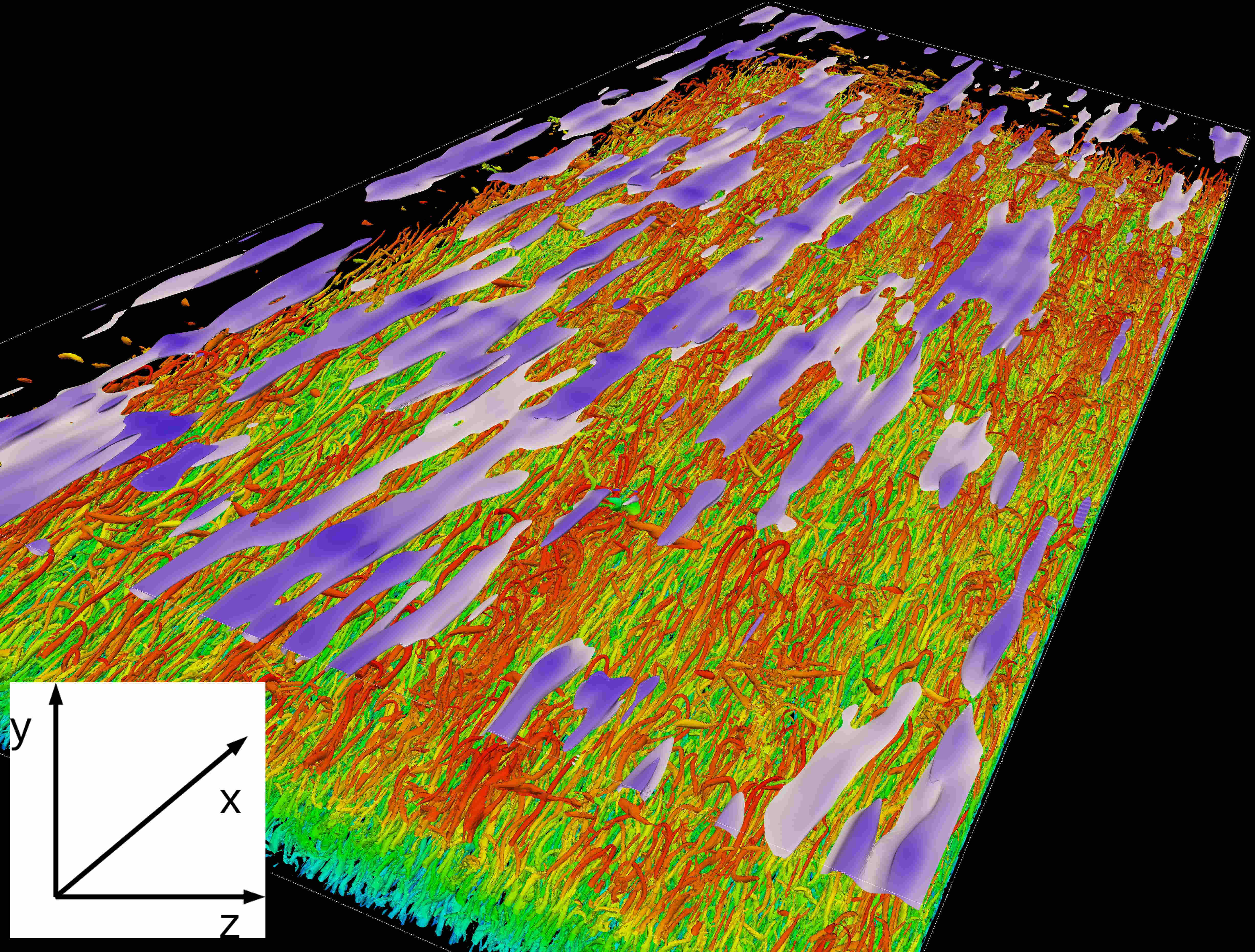}
\put(0.2,3.5){(d)}
%
\caption{Visualizations of the vortical structures on the HTS, with
instabilities and turbulent bursts on the WTS at
({\it a}) $t=7233$,
({\it b}) $t=7267$,
({\it d}) $t=7367$.
({\it c}) shows the
secondary instability at $t=7267$ in a wall-parallel plane.
Flow is from lower
left corner to upper right corner and the bottom side corresponds
to the HTS in ({\it a,b,d}).}
\label{vis}
\end{figure}

\begin{figure}
\setlength{\fwidth}{0.34\linewidth}
\setlength{\unitlength}{\fwidth}
\setlength{\unitlength}{1cm}
\includegraphics[width=42mm]{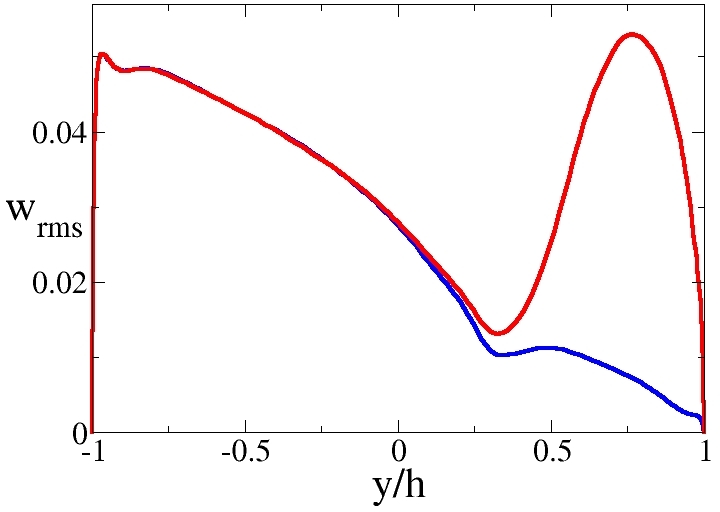}
\put(-1.2,2.6){(a)}
\hskip2mm
\includegraphics[width=42mm]{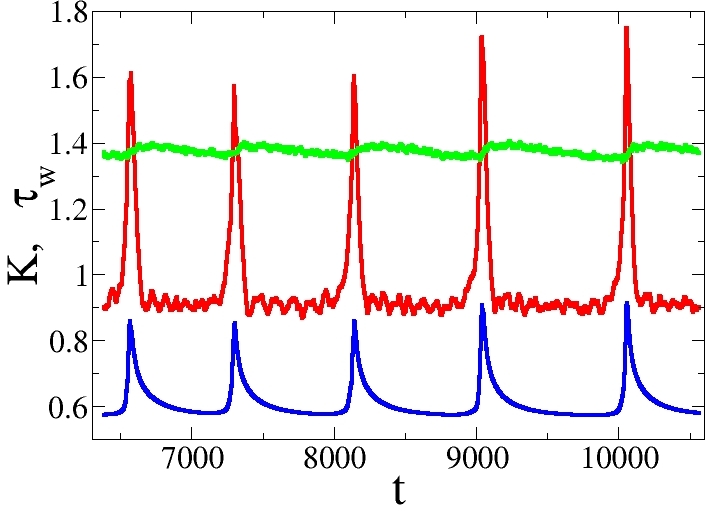}
\put(-1.2,2.6){(b)}
\caption{({\it a}) 
Root-mean-square of the spanwise velocity fluctuations
during periods without (\textcolor{blue}{---}) and 
with bursts (\textcolor{red}{---}).
({\it b}) 
Time series of the plane averaged wall shear stress on the HTS
(\textcolor{green}{---}) and WTS (\textcolor{blue}{---}), and 
volume integrated
turbulent kinetic energy (\textcolor{red}{---}).} 
\label{times}
\end{figure}

We focus now on cyclic turbulent bursts in RPF at 
$R=20000$ and $\Omega=1.2$.
Here $y \lesssim 0.25$ and $y \gtrsim 0.25$ correspond to 
the HTS and WTS respectively.
Vortical structures seen in Fig.~\ref{vis} illustrate
the vigorous turbulence on the HTS
with fluctuations $\sim 5\%$. 
Turbulent fluctuations on the WTS are less intense
due to the damping effect of rotation yet they are still significant in amplitude.
About 200 time units before a turbulent burst occurs,
a steadily growing plane wave 
with streamwise and spanwise wavenumbers 
$\alpha=2 \pi /\lambda_x =1$ ($\lambda_x$ is 
the streamwise wavelength) and $\beta=0$, respectively, 
and phase speed $\approx 0.2$,
appears on the WTS
(Fig.~\ref{vis}a).  
As the wave amplitude becomes large
the wave starts to bend owing to a secondary instability 
akin to H-type boundary layer transition \cite{herbert}
producing a typical staggered pattern of $\Lambda$-shaped vortices
(Fig.~\ref{vis}b,c).
The wave then breaks down 
into a burst of small-scale turbulence on the WTS
while turbulence on the HTS remains unaltered
(Fig.~\ref{times}a).
This intense turbulence on the WTS is damped by rotation 
and decays within 100 time units until the intensity reaches its initial level again
(Fig.~\ref{vis}d), whereafter
the plane wave starts to grow again.
This process leads to a self-sustained cycle
of intense turbulence bursts with 
sharp peaks in both turbulent kinetic energy
and wall shear stress on the WTS
(Fig.~\ref{times}b).
The cycle period $\sim 1000$ 
being much longer than typical turbulent 
time scales $\sim 1$ rules out a {\em direct}
forcing by the turbulence fluctuations.
The total turbulent kinetic energy grows by
$\approx 80\%$ during a burst, $\approx 20\%$ of which 
corresponds to the planar
$(\alpha = 1, \beta=0)$-wave.

This wave 
bears similarities with Tollmien-Schlichting (TS) waves 
produced via a linear instability of laminar Poiseuille flow, albeit there the
eigenfunction is symmetric and has a large amplitude on both channel sides.
To assess whether 
the cyclic bursts in RPF are caused by a similar instability,
we have carried out a linear stability analysis
of a specific base flow. The chosen base flow is incompressible and contains only
the streamwise component $U(y)$ of the velocity field spatially-averaged in the homogeneous
$x,z$-directions (the other two  components have zero mean).
Following standard procedures 
the linear equations for the 
wave-like perturbations $\vetu' = {\hat \vetu}(y) 
\exp [ i(\alpha x + \beta z - \omega t) ]$ 
with real wavenumbers $\alpha,\beta$
and complex frequency $\omega$ are derived,
yielding an eigenvalue problem for $\omega$.
Modes with $\beta=0$, such as the plane 
wave in Fig.~\ref{vis}, are unaffected 
by rotation, in contrast to 
modes with $\beta \neq 0$.
We only consider
the dominant $\beta=0$-modes whose
evolution, governed by the
Orr-Sommerfeld equations, depends on $R$
and the base flow $U(y)$ but not on $\Omega$.

\begin{figure}
\setlength{\fwidth}{0.34\linewidth}
\setlength{\unitlength}{\fwidth}
\setlength{\unitlength}{1cm}
\includegraphics[width=50mm]{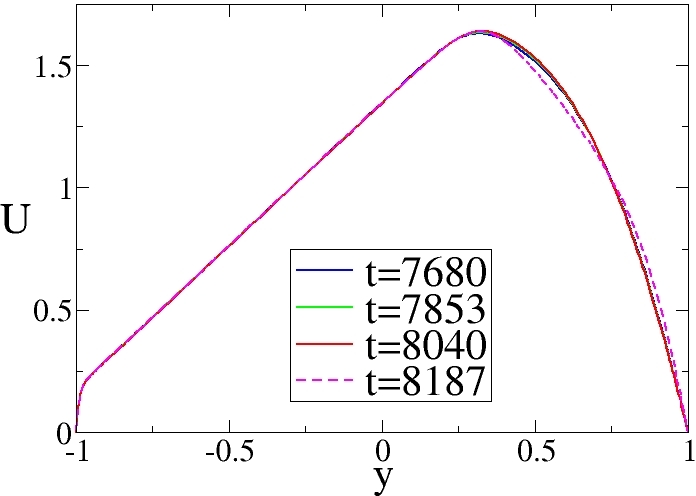}
\put(0.2,3.2){(a)}
\vskip2mm
\includegraphics[width=50mm]{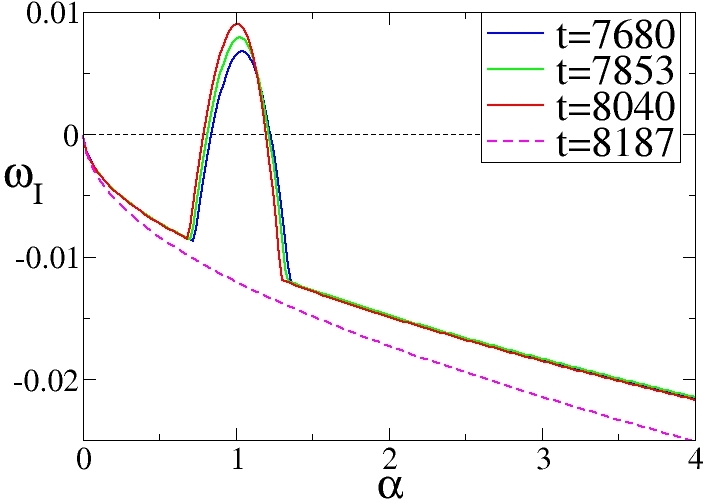}
\put(0.2,3.2){(b)}
\caption{({\it a}) Mean velocity profiles 
(WTS corresponds to $y \gtrsim 0.25$)
and ({\it b}) growth rates (imaginary part of $\omega$) versus $\alpha$ 
of $\beta=0$-modes 
at different times.}
\label{growth}
\end{figure}

Fig.~\ref{growth}a shows that this base flow hardly evolves
over the period $t=7680-8040$ before the burst, 
but after the burst at $t\approx 8140$
it changes on the WTS. 
We infer that
the turbulence has a limited {\em direct} influence on the perturbation
since its time and length scale are much smaller.
Fig.~\ref{growth}b shows the growth rate of the most unstable eigenvalue 
of the $\beta=0$-modes at various times.
Perturbations with $(\alpha \approx 1, \beta=0)$ 
have a single unstable eigenvalue
prior to the burst at $t \approx 8140$ and 
are thus linearly unstable, 
while over a short period after the burst all $\beta=0$-modes are stable
and decay according to linear analysis.
The same pattern is found before and after 
every burst, confirming the robustness
of the results.
Including a turbulent viscosity in the stability analysis
\cite{alamo} lowers the growth rate by about 10\%, yet it does not affect
the results significantly, suggesting that turbulence has a 
limited direct influence despite its intensity.
We can thus infer that rotation modifies the mean flow $U(y)$ so that it becomes receptive
to a linear instability;
non-rotating plane turbulent channel flow has a linearly stable base flow
$U(y)$ and does not show large-scale instabilities.

\begin{figure}
\setlength{\fwidth}{0.34\linewidth}
\setlength{\unitlength}{\fwidth}
\setlength{\unitlength}{1cm}
\includegraphics[width=42mm]{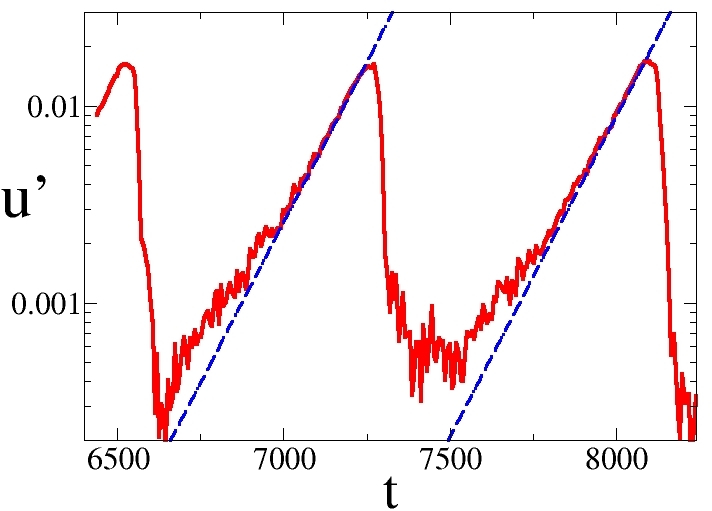}
\put(-1.0,2.6){(a)}
\hskip2mm
\includegraphics[width=42mm]{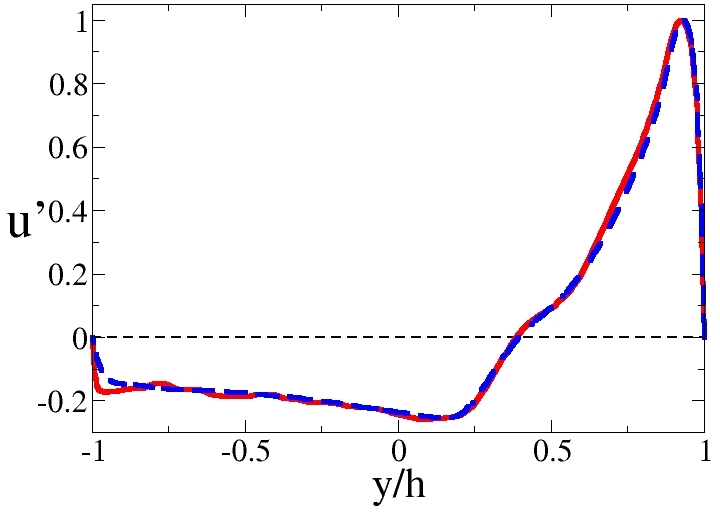}
\put(-1.0,2.6){(b)}
%
\caption{({\it a}) Amplitude $u'$ and ({\it b}) profile at $t=8040$
of the $(\alpha=1, \beta=0)$-mode 
in the simulation (\textcolor{red}{---}) compared to the growth rate and eigenfunction
predicted by the linear stability analysis (\textcolor{blue}{$---$}).}
\label{amp}
\end{figure}

Fig.~\ref{amp}a shows the exponential growth of the $(\alpha = 1, \beta=0)$-mode, with a mean growth rate of $\approx 7.5 \times 10^{-3}$ as predicted by the stability analysis, compared to its time evolution extracted from the simulation.
The amplitude $u'$ of the mode, corresponding to 
the plane wave in Fig.~\ref{vis},
grows by two orders of magnitude.
The wave amplitude needs 307 time units to 
grow by one decade, which explains the long intervals between bursts.
Variations in the bursting period are due to
the stochastic nature of the background fluctuations. 
The eigenfunction 
(Fig.~\ref{amp}b) 
and frequency
obtained from linear analysis also match simulation results, proving
that a linear instability with a large amplitude on the WTS 
produces the exponentially growing plane wave in RPF
and is the principal driving mechanism for the cyclic bursts.
This process is self-sustained since, 
besides the driving pressure gradient, no external trigger
is needed to maintain it.
In the simulation the pressure force was varied 
to keep the mass flow rate constant,
but a constant pressure force produces 
essentially the same cyclic bursts.
%

We may speculate on what happens in currently 
out-of-reach simulations with extended domains. 
The linear instability would still occur in such domains, 
but the competition between different wavelengths and phases
could lead to non-uniform or even localized bursts and hence
spatio-temporal intermittency.

RPF simulations to be reported in a forthcoming study, 
mapping out an extensive parameter range,
demonstrate that cyclic bursts also happen 
at higher $\Omega$ and higher or lower $R$ 
until a critical $R_c=3848$ when TS waves become 
stable in Poiseuille flow according to linear theory. 
%
%
However, at higher $\Omega$ and lower $R$ 
turbulence is less intense and RPF can even partly or completely relaminarize \cite{olof}; 
instabilities in a turbulent environment are thus merely found at higher $R$.
Relaminarization and extreme amplitude states at low $R$
have recently been studied \cite{wallin}.

Other external forces or conditions can also alter 
turbulence and mean velocity profiles,
suggesting that instabilities can happen in 
other hydrodynamical systems.  
Cyclic bursts of turbulence caused by a 
linearly unstable TS wave have indeed been observed in
low magnetic Reynolds number
 Poiseuille flow with a steady spanwise magnetic 
field at $R=5333$ \cite{boeck,dey}. 
Between the bursts the flow is fully laminar 
since the Lorentz force suppressed turbulence. 
Our study shows that the absence of turbulence
is not a prerequisite for a linear instability. 
Indeed, we conducted magnetic simulations 
at $R=20000$, where the flow does not fully relaminarize, and identified cyclic bursts for strong magnetic fields similar to RPF.
Unlike in
\cite{boeck}, no noise is needed to sustain the cycle 
since the flow is always subject to turbulent fluctuations.
A difference to RPF is the statistical symmetry of the magnetic case about the centreline.

To summarize, 
rotation changes the flow in strongly turbulent RPF in such a way that it becomes
receptive to a linear instability, 
despite the strong turbulence,
and a simple reduced model can describe this
driving instability. 
The turbulence results from deterministic dynamics 
as opposed to parametrizable external noise,
and has a wide range of temporal and spatial scales.
We are able to describe and understand all phases leading to the
recurrent intense low-frequency bursts. 
This case strongly suggests
a low-dimensional dynamic system embedded 
in an environment with strong fluctuations.
Low-dimensional bursting dynamics are also found 
in accretion disks \cite{ac} and tokamaks \cite{strugarek}, where
the full interaction between large-scale modes
and small-scale fluctuations deserves further study. The present study offers some important insights; 
environments with inhomogeneous strong fluctuations can 
support cyclic instabilities of a large-scale mode.
The simple and well-defined 
RPF with its multifaceted physics 
is a valuable alternative to other flows featuring shear and rotation
such as Taylor-Couette and von 
K{\'a}rm{\'a}n flows.

\begin{acknowledgments}
We acknowledge PRACE for awarding us via the REFIT project
access to resource {\em Jugene}
at the J{\"u}lich Supercomputing Centre in Germany.
Computational resources at PDC were made available by SNIC.
We thank Liang Wei for producing some of the visualizations.
The project was supported by the Swedish Research Council through Grants
No. 621-2010-4147 and No. 621-2013-5784.
\end{acknowledgments}



\end{document}